\documentclass[aps,prl,reprint,groupedaddress]{revtex4-1}
\usepackage{multirow}
\usepackage{graphicx}
\newcommand{\sro}{$\textrm{Sr}_{2}\textrm{RuO}_{4}$}

\begin{document} 


\title{Spin-orbit induced mixed-parity pairing in Sr$_2$RuO$_4$:
a quantum many-body calculation} 

\author{J.J. Deisz}
\email{john.deisz@uni.edu}
\author{T.E. Kidd}
\affiliation{Department of Physics, University of Northern Iowa,
Cedar Falls, Iowa 50614, USA}

\date{\today}

\begin{abstract}
The unusual superconducting state in Sr$_2$RuO$_4$
has long been viewed
as being analogous to a superfluid state in liquid $^3$He. 
Nevertheless, calculations based on this odd-parity state are
presently unable to completely reconcile the
properties of Sr$_2$RuO$_4$.
Using a self-consistent
quantum many-body scheme 
that employs realistic parameters, 
we are able to model several signature properties of
the normal
and 
superconducting states of Sr$_2$RuO$_4$.
We find that the dominant component 
of the model superconducting state is of even parity and closely
related to superconducting state 
for the high-$T_c$ cuprates although a
smaller odd-parity component is induced by spin-orbit coupling.
This mixed pairing state gives a more complete 
representation of the complex phenomena measured in Sr$_2$RuO$_4$.
\end{abstract}

\pacs{74.20.Rp,74.70.Pq}

\maketitle

The unusual electronic properties
of the layered material Sr$_2$RuO$_4$ remain
of significant interest eight years after Mackenzie and Maeno's 
review \cite{RevModPhys.75.657} of an already extensive literature.
Experiments show
a quasi two-dimensional Fermi liquid 
state for $T < 100\textrm{K}$ with an unconventional superconducting state 
emerging at $T_c = 1.5\textrm{K}$. 
A qualitative description of the normal state Fermi liquid
has emerged through band structure calculations.
Three quasiparticle 
bands, $\alpha$, $\beta$ and $\gamma$,
originating from $d_{xz}$, $d_{yz}$
and $d_{xy}$ orbitals of tetragonally-coordinated Ru ions,
cross the Fermi surface \cite{PhysRevB.51.1385,Mackenzie1996510}.
The nature of the superconducting state is less well
established, but the lack of in-plane Knight shifts
for $T < T_c$ \cite{knight-in-plane} and the presence of
finite signals in symmetry sensitive probes such as 
the polar Kerr \cite{PhysRevLett.97.167002} or
the Josephson effects \cite{Nelson12112004,Kidwingira24112006}
point to spin-triplet, odd-parity pairing and, more specifically,
to a chiral pairing state with zero total pair spin
perpendicular to the planes, \textit{i.e}
$S_z = 0$.
Nonetheless, other experiments are not so well-explained
by such a pairing state \cite{0953-8984-21-16-164210}: 
the specific heat for $T << T_c$
\cite{springerlink:10.1023/A:1022551313401}
shows evidence of nodes or minima
in the superconducting gap function, 
but a
chiral $p$-wave state is nodeless;
the out-of-plane Knight shift \cite{PhysRevLett.93.167004}
is weakly temperature dependent below $T_c$;
and 
edge currents are not detectable
at a level that is two orders of magnitude below the expected 
value \cite{PhysRevB.76.014526}.
More recently, signatures of
half-quantum vortices below $T_c$ were explained in
terms of equal spin pairing, \textit{i.e.}
$S_z \neq 0$ \cite{Jang14012011}.
Thus, the nature of the bulk superconducting state is still 
not resolved completely and
it is essential to test current working assumptions
with microscopic calculations that are as detailed as possible.

Our calculation starts with  
a tight-binding representation of
an LDA band-structure for Sr$_2$RuO$_4$ as developed by 
Pavarini and Mazin \cite{PhysRevB.74.035115}.
Low energy electronic states are represented
with a basis consisting of a $d_{xy}$, $d_{xz}$ and
$d_{yz}$ orbital for each Ru ion
and these are 2/3 filled on average. 
Ru ions are strongly coupled within planes
where they form a square lattice
of side $a = 3.87$\AA.
Weak interplanar couplings corresponding
to the $z$ direction are ignored.
For the Ru atomic spin-orbit interaction, $\lambda ~ \vec{s}\cdot \vec{l}$,
we use 
$\lambda = 70\textrm{ meV}$ \cite{0295-5075-49-4-473}.
The interaction of atomic spin and orbital moments and a
uniform external field
is given by $-\mu_B (g_s \vec{s} + \vec{l})\cdot \vec{B}$.

Dynamical correlation effects are modeled on the basis of
an atomically local bare interaction
with three parameters; $U$ and $U^{\prime}$ for
inter- and intra-orbital Coulomb repulsion
and $J$ for interorbital exchange; and the fluctuation
exchange approximation (FLEX).
Here we will only consider fixed ratios of $U^{\prime}/U = 2/3$ and
$J/U = 1/6$ that are typically used in these 
calculations \cite{PhysRevLett.106.096401}; variations
in these ratios have little impact on general features of our results.
Having fixed the ratios we find that 
$U < 1 \textrm{ eV}$ is reasonably consistent with
the observed $T_c$ = 1.5K, but
we will in some instances use a somewhat larger value 
of $U = 1.2\textrm{ eV}$
as the higher $T_c$ that results makes superconducting state calculations
more feasible.
To make the scheme computationally feasible,
the dynamical cluster approximation \cite{PhysRevB.58.R7475} 
is used to reduce
dynamical correlations to a range of $L_c << L$ where $L \sim 100 a$
is the physical lattice dimension and we numercaly demonstrate that our
results well-describe the large $L_c$ limit.
Translationally invariant, but otherwise random initial configurations
are used 
to seed
the self-consistent procedure which then terminates when a stable 
state is
produced. 

\begin{figure}[!h]
\begin{center}
\includegraphics[scale=0.5]{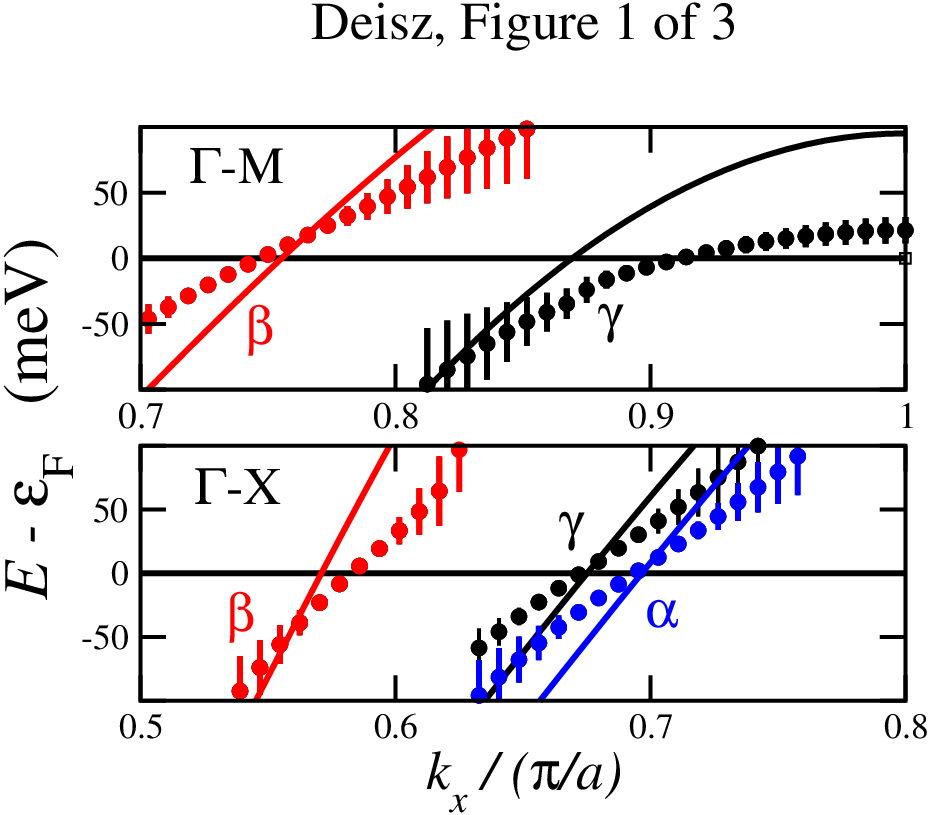}
\end{center}
\caption{\label{normalqp}
(Color online) 
Comparison of low energy band structure derived from FLEX calculations 
(symbols) and LDA (solid lines) along the indicated symmetry directions. 
The vertical lines through each point correspond to the inverse 
lifetime of the state. There is a significant renormalization of each 
band, with the $\gamma$ band showing the strongest renormalization 
along $\Gamma$-M.
}
\end{figure}

Figure~\ref{normalqp} compares the renormalized normal state ($T=60\textrm{K}$)
electronic structure of Sr$_2$RuO$_4$ as determined by FLEX
calculations for $U = 0.9\textrm{eV}$ 
with bare band
dispersions derived from LDA \cite{PhysRevB.74.035115}.
The FLEX band dispersions
have significantly smaller slopes at the Fermi level than
unrenormalized LDA counterparts. These indicate a
significant mass enhancement for each band that is 
expected from
specific heat \cite{maeno} and dHvA \cite{Bergemann2001371} measurements. 
The imaginary part of the self energy for each band is proportional to
$(E - \varepsilon_F)^2$ as 
is expected for a Fermi liquid.  

As the
calculations can be used to determine the electronic structure
with both energy and momentum resolution, it is natural to
present a comparison to angle resolved photoemission
experiments. Fortunately, once the problem of a surface
reconstruction was resolved \cite{PhysRevLett.85.5194},  
there have been a plethora of
high quality photoemission experiments that probed the bulk
normal state electronic structure of 
\sro \cite{PhysRevB.72.205114,PhysRevLett.94.107003,
PhysRevB.72.104514,PhysRevLett.93.117005}. 
For the $\beta$ 
and $\gamma$ band crossings along $\Gamma$-M, 
the calculated dynamical renormalizations of the 
Fermi velocity, 
$v_{F,LDA}/v_{F,FLEX}$ where $v_F= dE/dk|_{k=k_F}$,
are 1.9 and 3.4 which compare well
to the experimental values
of (approximately) 1.7 and 3.3 \cite{PhysRevB.72.104514}.

A more striking renormalization
is the shift
of the van Hove singularity
in the $\gamma$-band
at the M point 
from 90~meV to 20~meV in going from the LDA to FLEX results.
This feature is clearly apparent in photoemission
experiments utilizing either thermally occupied states or doping
\cite{PhysRevLett.99.187001}. This confirms that FLEX is capturing
non-trivial normal-state correlation effects that could be key
precursors to the superconducting transition.
We note that spin-orbit coupling
leads to non-negligible spin and orbital
mixing for $\gamma$-band quasiparticles
at $\epsilon_F$ along $\Gamma$-M with
that $d_{xz}$ orbitals contributing about 15\% of the quasiparticle weight.
Since the spin-orbit induced mixing of 
$d_{xy}$ and $d_{xz}$ orbitals 
involves a spin flip, i.e. $l^+ s^-$, 
these quasiparticles
have a wavefunctions of the form 
$|\psi^{qp}> \sim c_1 |d_{xy},\uparrow> + c_2 |d_{xz},\downarrow>$, 
\textit{i.e.} they are not pure spin-eigenstates.

Superconductivity 
is associated with 
the stability of finite
anomalous Green's function amplitudes, 
$F_{\nu\sigma\nu^{\prime}\sigma^{\prime}}(\vec{R},\tau)
\equiv -{\cal T}_{\tau} < c^{\dagger}_{\nu\sigma}(\vec{R},\tau)\,
c^{\dagger}_{\nu^{\prime}\sigma^{\prime}}(\vec{0},{0})>.$
As an order parameter amplitude
for the superconducting state we take
$m_p^2(T) = \sum_{\nu\sigma\nu^{\prime}\sigma^{\prime}\vec{R}}
| F_{\nu\sigma\nu^{\prime}\sigma^{\prime}}(\vec{R},\tau=0) |^2$
and fit the temperature dependence of $m_p(T)$ to
a power-law function, $a_o|T-a_2|^{a_1}$ to estimate
$T_c = a_2$. Results for $T_c$ versus $U$, $\lambda$ and
$L_c$ are shown in Table~1.
While the numerical parameter $L_c$ does affect quantitative results,
qualitative features are reasonably well described 
with $L_c = 4a$.
With spin-orbit coupling ($\lambda = 0.07$)
we find that $T_c \sim 20-30\textrm{K}$ for
$U$=1.2~eV and $T_c \lesssim 10\textrm{K}$ for $U$ = 0.9~eV.
Thus, $U \lesssim 0.9$~eV is most consistent with the observed
$T_c$, unless the spin-orbit interaction
for Sr$_2$RuO$_4$ in actuality is much larger than 0.07~eV.
Recall that $U$=0.9~eV leads to quantitative agreement with
photoemission results as well.
\begin{table}[!h] 
\begin{center}
\begin{tabular}{| c | c | c || c | c | c | c | c |}
\hline
\multicolumn{3}{|c||}{model parameters} & \multicolumn{5}{|c|}{calculated values} \\
\hline
$U$ (eV) & $\lambda$ (eV) & $L_c/a$ & $T_c$ (K)&  $P^{(0,0)}$ & $P^{(1,1)}$ & $P^{(1,0)}$ 
& $P^{(1,-1)}$ \\
\hline
1.2  & 0.00 & 4 & 39 & 1.0 & 0 & 0 & 0 \\
1.2 & 0.00 & 8 & 29 & 1.0 & 0 & 0 & 0 \\
1.2 & 0.02 & 4 & 39 & 0.986 & 0.0070 & 0.000021 & 0.0070 \\
1.2  & 0.07 & 4 & 35 & 0.848 & 0.075 & 0.0019 & 0.075 \\
1.2 & 0.07 & 8 & 22 & 0.785 & 0.106 & 0.0018 & 0.106 \\
1.2 & 0.07 & 16 & 29 & 0.795 & 0.101 & 0.0021 & 0.101 \\
0.9 & 0.00 & 4 & 15 & 1.0 & 0 & 0 & 0 \\
0.9 & 0.07 & 4 & 6 & 0.762 & 0.118 & 0.0023 & 0.118 \\
0.9 & 0.07 & 8 & 6 & 0.713 & 0.142 & 0.0017 & 0.142\\
\hline
\end{tabular}
\end{center}
\caption{The transition
temperature, $T_c$, and pair wave function weights by spin, $P^{(S,S_z)}$,
as a function of the
Coulomb repulsion strength ($U$), and spin-orbit coupling constant ($\lambda$)
and dynamical cluster size ($L_c$). 
In the absence of spin-orbit coupling ($\lambda=0$), the 
pairing state is a pure singlet.
Spin-orbit coupling induces triplet components to the
pairing state, primarily between electrons with equal $z$-components
of spin, $S_z = \pm 1$.}
\end{table}

To examine the spatial, orbital and spin dependence of the pairing
state, we define a normalized pair wave function 
and decompose it into eigenfunctions of total spin, $(S, S_z)$,
\begin{displaymath}
\Psi_{\nu\sigma\nu^{\prime}\sigma^{\prime}}(\vec{R})
= \frac{1}{m_p} F_{\nu\sigma\nu^{\prime}\sigma^{\prime}}(\vec{R},0) =
\sum_{S,S_z} \Psi^{(S,S_z)}_{\nu\nu^{\prime}}(\vec{R})\chi^{(S,S_z)}_{\sigma\sigma^{\prime}}
\end{displaymath}
from which we obtain a fractional weight by total spin,
$P^{(S, S_z)}$, by taking the complex squares 
of $ \Psi^{(S,S_z)}_{\nu\nu^{\prime}}(\vec{R})$ and tracing over orbital indices
and displacements.
In Table~1 we show results for 
these weights as a function
of $U$, $\lambda$ and $L_c$ evaluated at $T \simeq T_c$.
These results show a pure singlet
pairing state when  
spin-orbit coupling is absent, \textit{i.e} $\lambda=0$. 
Further examination of the orbital
and spatial properties  
of $\Psi^{(0,0)}_{\nu\nu^{\prime}}(\vec{R})$ show that this singlet
state has $d_{x^2-y^2}$ symmetry with respect to rotations
in the plane,  \textit{i.e.}
$\Psi_{xy;xy}^{(0,0)}(a\hat{y}) = -\Psi_{xy;xy}^{(0,0)}(a\hat{x})$,
and that the only significant orbital contributions come from
the $d_{xy}$-orbitals with the very small contributions from 
$d_{xz}$ and $d_{yz}$ orbitals vanishing completely
in the limit $J \to 0$.
Consequently, for
$\lambda =0$ 
we find that the superconducting state leaves
the $\alpha$ and $\beta$ bands ungapped or very nearly so
and the $\gamma$-band gap has nodes typical of a $d_{x^2-y^2}$
pairing state.

A finite spin-orbit coupling constant, $\lambda \neq 0$, 
finite
triplet pairing components.
The largest amplitudes for triplet pairing occur for equal
spin pairs, \textit{i.e.} $S_z = \pm 1$. 
Further examination of 
the wavefunctions for $S_z = \pm 1$ 
reveals that they describe
interorbital pairing between an electron in the $d_{xy}$ orbital 
and an electron in either of the $d_{xz}$ of $d_{yz}$ orbitals. 
Also we find 
$\pm \pi/2$ phase shifts between the two 
types of interorbital terms in this wavefunction
with opposite relative signs for the up and down 
spin pairs, \textit{i.e} 
$\Psi^{(1,1)}_{xy,yz}(a\hat{y}) =
 e^{-i \pi/2} \Psi^{(1,1)}_{xy,xz}(a \hat{x} )$
and $\Psi^{(1,-1)}_{xy,yz}(a\hat{y}) =
 e^{+i \pi/2} \Psi^{(1,-1)}_{xy,xz}(a \hat{x} )$.
Odd-parity states of this form are ``topological,''
time-reversal invariant
superconducting states and their unique properties
have recently 
been described by Qi, \textit{et al.} 
\cite{PhysRevLett.102.187001}.
Tada \textit{et al.} \cite{1367-2630-11-5-055070} have performed
model calculations that point to the possibility of generating
such a state at the interface via spin-orbit coupling although they suggest
that weaker spin-orbit effects lead to
a chiral $p$-wave state in the bulk.
Note that 
for such a state up and down spin equal-spin
pairs have opposite chirality, so no net charge current is expected
on the edges, a result that is consistent
with experiment \cite{PhysRevB.76.014526}. 
We note that for fixed values of $U$ and $\lambda$, the relative contribution
of singlet and triplet pairing terms 
is weakly dependent on the numerical parameter $L_c$ with little change 
observed in going from $L_c/a =8$ to 16.

\begin{figure}[!h]
\begin{center}
\includegraphics[scale=0.5]{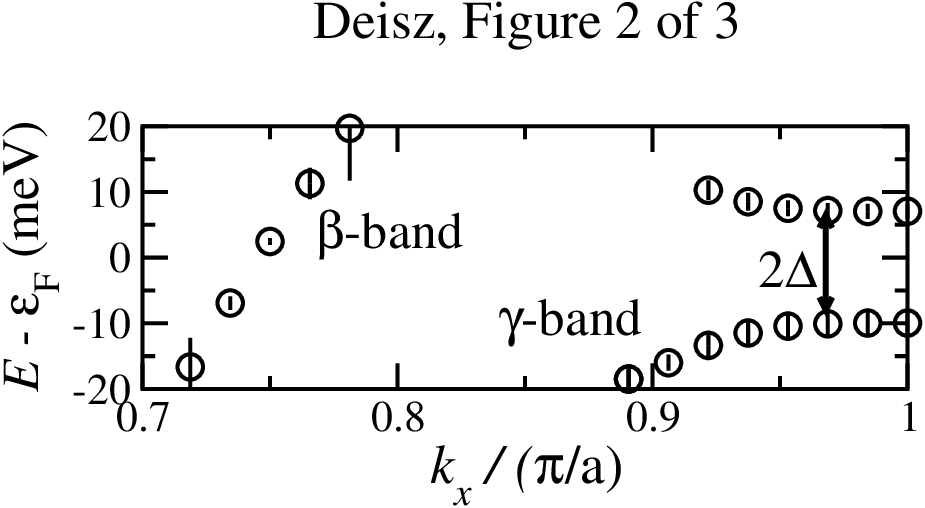}
\end{center}
\caption{\label{scqp}
Quasiparticle spectrum along the $\Gamma$-M 
direction for $T < T_c$. The usual
signature of a superconducting gap appears in the $\gamma$-band.
There is no
evidence for a smaller gap forming in the $\beta$-band along this cut
to within the momentum resolution of the calculation.
}
\end{figure}

In Figure~\ref{scqp}, we show the quasiparticle excitation
spectrum for $T \sim 0.5 T_c$ along the $\Gamma$-M cut
evaluated with $U=1.2$~eV, $\lambda = 0.07$~eV and
$L_c = 8a$.
The characteristic excitations
on both sides of the superconducting gap are evident in the
$\gamma$-band near $k_x = \pi/a$. 
No extra gap is readily apparent in 
either the $\alpha$ or $\beta$ bands, 
or in any of the bands along the zone diagonal.
This is 
the structure observed for pure $d$-wave pairing when 
$\lambda = 0$ and while new, but smaller gap features are expected
with the triplet component, they are not readily apparent
given the momentum/energy resolution (as indicated by the plotted points)
in this calculation.
Recall that 
nodal structure in the superconducting gap observed
in these FLEX results is consistent with observations
for the specific heat
superconducting Sr$_2$RuO$_4$ \cite{springerlink:10.1023/A:1022551313401}
whereas a chiral $p$-wave state is expected to be nodeless. However,
it remains to be demonstrated that a superconducting gap function
of the form shown here is consistent with all probes that 
are sensitive to the quasiparticle excitation spectrum below $T_c$,
all of which are consistent with a nodal strucutre, but as of yet,
of undetermined form.

\begin{figure}[!h]
\begin{center}
\includegraphics[scale=0.4]{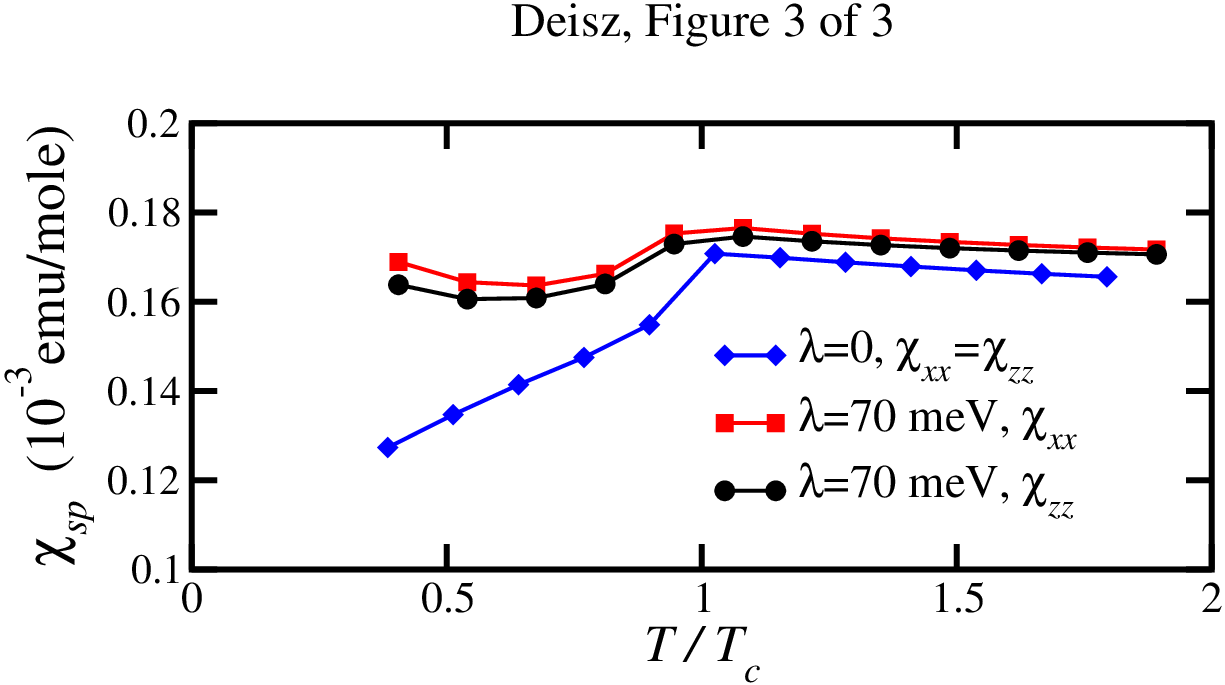}
\end{center}
\caption{\label{chi}
Temperature dependence
of the Pauli spin susceptibility, $\chi_{sp}$,
for zero and finite spin-orbit coupling constant, $\lambda$.
For $\lambda=0$ the equal in-plane and out-of-plane components 
are suppressed by the superconducting transition
In contrast, for $\lambda = 70\textrm{ meV}$, both the in-plane
and out-of-plane components are suppressed by less than 10\% for
$T < T_c$. 
}
\end{figure}

Next we consider the temperature dependence of the
Pauli spin susceptibility.
Measurements of the 
Knight shift 
for $0 < T < T_c$
suggest a
weak variation with temperature for both 
$\chi_{xx}$ and $\chi_{zz}$, the in-plane and out-of-plane components of
the uniform 
Pauli susceptibility \cite{knight-in-plane,PhysRevLett.93.167004}.
Calculated susceptibility results 
for $U=1.2$~eV and $L_c = 4a$ as a function
of reduced temperature, $T/T_c$,
are displayed in Figure~\ref{chi}.
For $\lambda=0$ we find that $\chi_{xx} = \chi_{zz}$ is
suppressed by about 1/3 below $T_c$, but the curves do not approach
zero as $T \to 0$ as the $\alpha$-band and $\beta$-band Fermi surfaces
remain ungapped.
In the mixed pairing state, for $\lambda = 0.07$~eV,
$\chi_{xx}$ and $\chi_{zz}$ are no longer equal, but
each has a weak temperature dependence below $T_c$ with
changes of less than 10\% compared to $T \sim T_c$.
Thus, the qualitative behavior
of the temperature-dependent
spin susceptibility
for this mixed pairing state apparently is in good agreement
with Knight shift data.
Further,
these results suggest that the small Knight shifts
in Sr$_2$RuO$_4$ are caused by the mixed-spin nature of 
quasiparticle states, a general mechanism
for small Knight shifts described by 
Anderson \cite{PhysRevLett.3.325},
and not by a bulk $p$-wave symmetry pairing state.  

We note, though, that the overall
scale for the calculated susceptibilities are 
a factor of about four too small
in comparison to experiment \cite{maeno}. The lack of quantitative
accuracy in this instance
is most likely a failure
of FLEX as the approximation has been shown to underestimate
the large enhancements of the magnetic susceptibility 
that are expected in the vicinity
of a magnetic phase transition
for the special case of the
single-band Hubbard model \cite{PhysRevB.43.8044}.

Thus, the mixed pairing superconducting state is consistent 
with major experimental features of Sr$_2$RuO$_4$ in 
both the superconducting and normal states. 
However, the time-reversal symmetry preserving nature of this
mixed-parity state
places these results
at odds with the interpretation
of polar Kerr and tunneling experiments 
\cite{PhysRevLett.97.167002,Kidwingira24112006}.
We note that such an interpretation of the tunneling data
requires the presence
of superconducting domains of size $\lesssim$~1$\mu$m,
which is too small to account for a finite signal in the infrared
frequencies used in the polar Kerr effect experiments
\cite{0953-8984-21-16-164210}.
Also the existence of a finite polar Kerr effect at the
pseudogap temperature, $T^*$, in the cuprate superconductors
suggests that this probe is
sensitive to correlations other than those directly related to
a time-reversal symmetry breaking superconducting
state \cite{He25032011}.
It remains to 
be seen whether the correlation between experiment and 
calculations can be improved by accounting for additional 
details such as vortices, defects, impurities, 
surface effects, domains and other perturbations, 
all of which have been found to be necessary for 
enabling the theory for a pure $p$-wave superconducting ground state 
to be consistent with the wide range of experimental studies of the system.

The results shown here are supported by earlier theoretical work. For
example, Kuwabara and Ogata \cite{PhysRevLett.85.4586} showed that
fluctuations may favor a $d_{x^2-y^2}$ pairing given certain
parameters in an effective one-band model for \sro.  In the course of
developing a theory for a $p$-wave superconducting state for \sro, it
has long been understood that spin-orbit coupling might play a role in
stabilizing one of the possible $p$-wave  pairing states
\cite{0295-5075-49-4-473}, but less attention has been placed on the
impact of spin-orbit coupling on a nominally $d_{x^2-y^2}$ pairing
state.  However, at the conclusion of their study of spin-orbit
effects on the  band structure of Sr$_2$RuO$_4$  and Sr$_2$RhO$_4$,
Haverkort, \textit{et al.}  state ``In fact, singlet and triplet
states could be mixed, blurring the distinction between spin-singlet
and  spin triplet-pairing ... \cite{PhysRevLett.101.026406}.''  Indeed,
we find that when dynamical correlations \textit{and} spin-orbit
interactions are simultaneously taking into account using realistic
parameters for Sr$_2$RuO$_4$ such a mixed-parity bulk state is
realized.  As FLEX is an approximate many-body scheme its
identification of the superconducting state for the underlying
three-band model must be considered tentative until it is
computationally feasible to include terms that are missing in FLEX
(i.e. vertex corrections in the self-energy series) or perform a
quantum Monte Carlo analysis. Nonetheless, this FLEX analysis
represents is a methodological step forward as it incorporates
physical effects not included in earlier finite-order perturbative
studies \cite{JPSJ.69.3678,JPSJ.72.673}, namley (i) collective spin
fluctuations via an infinite set of ladder diagrams and  (ii)
dynamical quasiparticle renormalizations via self-consistency of the
electron Green’s function. As these effects are not small, it is
unsurprising that a different conclusion results with respect to the
pairing symmetry.


T.E.K acknowledges support from the Iowa Office 
of Energy Independence Award No. 09-IPF-11. J.J.D. as supported in part by a Professional Development 
Assignment from the University of Northern Iowa. 
J.J.D. thanks Paul Gray
of the University of Northern Iowa for 
computational support and R.R.P. Singh and
the Physics Department of the University of California, 
Davis for hosting him and many useful conversations 
during the time this work was initiated.


%
\end{document}